\documentclass[%
reprint,
superscriptaddress,
 amsmath,amssymb,
 aps,
floatfix,
]{revtex4-1}

\usepackage{graphicx}
\usepackage{dcolumn}
\usepackage{bm}

\usepackage[flushleft]{threeparttable} 

\usepackage{dsfont}
\usepackage{color,psfrag}

\usepackage{xcolor} 

\usepackage{soul}

\renewcommand\vec{\mathbf}

\newcommand{\mat}[1]{\mathbf{#1}} 

\newcommand{\tran}{^{\mathstrut\scriptscriptstyle\top}} 

\begin{document}

\title{Towards Exact Molecular Dynamics Simulations with Machine-Learned Force Fields}

\author{Stefan Chmiela}
 \affiliation{Machine Learning Group, Technische Universit\"at Berlin, 10587 Berlin, Germany}

\author{Huziel E. Sauceda}
 \affiliation{Fritz-Haber-Institut der Max-Planck-Gesellschaft, 14195 Berlin, Germany}

\author{Klaus-Robert M\"uller}%
\email{klaus-robert.mueller@tu-berlin.de}
\affiliation{Machine Learning Group, Technische Universit\"at Berlin, 10587 Berlin, Germany}
\affiliation{Department of Brain and Cognitive Engineering, Korea University, Anam-dong, Seongbuk-gu, Seoul 136-713, Korea}
\affiliation{Max Planck Institute for Informatics, Stuhlsatzenhausweg, 66123 Saarbr\"ucken, Germany}

\author{Alexandre Tkatchenko}
 \email{alexandre.tkatchenko@uni.lu}
\affiliation{Physics and Materials Science Research Unit, University of Luxembourg, L-1511 Luxembourg, Luxembourg}

\date{\today}
         
\begin{abstract}
Molecular dynamics (MD) simulations employing classical force fields constitute the cornerstone of contemporary atomistic modeling in chemistry, biology, and materials science. However, the predictive power of these simulations is only as good as the underlying interatomic potential. Classical potentials often fail to faithfully capture key quantum effects in molecules and materials. Here we enable the direct construction of flexible molecular force fields from high-level \emph{ab initio} calculations by incorporating spatial and temporal physical symmetries into a gradient-domain machine learning (sGDML) model in an automatic data-driven way. The developed sGDML approach faithfully reproduces global force fields at quantum-chemical CCSD(T) level of accuracy and allows converged molecular dynamics simulations with fully quantized electrons and nuclei. We present MD simulations, for flexible molecules with up to a few dozen atoms and provide insights into the dynamical behavior of these molecules. Our approach provides the key missing ingredient for achieving spectroscopic accuracy in molecular simulations.
\end{abstract}

\maketitle


\section{Introducion}
Molecular dynamics (MD) simulations within the Born-Oppenheimer (BO) approximation constitute the cornerstone of contemporary atomistic modeling. In fact, the 2013 Nobel Prize in Chemistry clearly highlighted the remarkable advances made by MD simulations in offering unprecedented insights into complex chemical and biological systems. However, one of the widely recognized and increasingly pressing issues in MD simulations is the lack of accuracy of underlying classical interatomic potentials, which hinders truly predictive modeling of dynamics and function of (bio)molecular systems.
\begin{figure*}[!p]
\centering
\includegraphics[width=\textwidth]{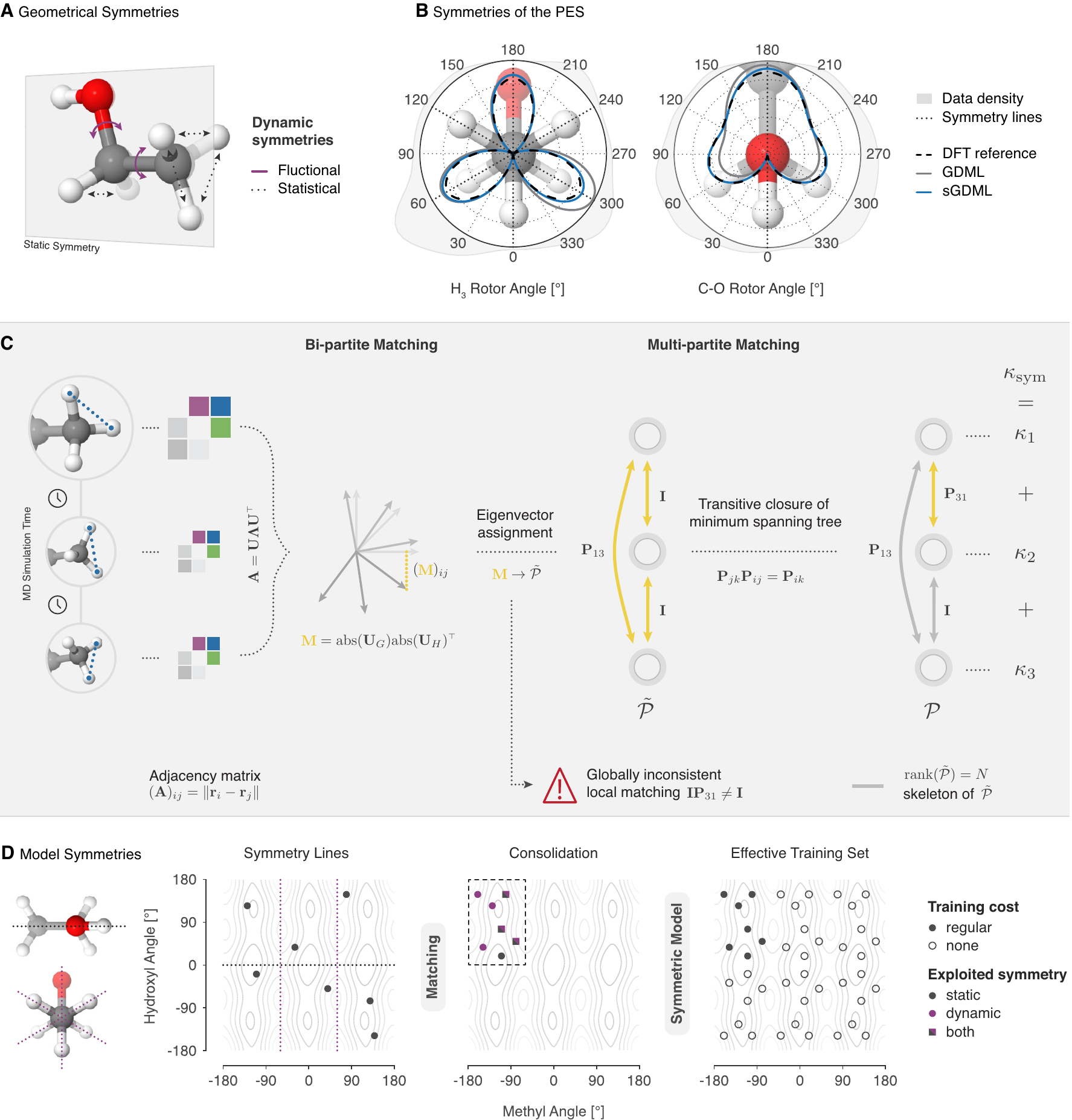}
\caption{Fully data-driven symmetry discovery. (A, B) Our multipartite matching algorithm recovers a globally consistent atom-atom assignment across the whole training set of molecular conformations, which directly enables the identification and reconstructive exploitation of relevant spatial and temporal physical symmetries of the molecular dynamics. (C) The global solution is obtained via synchronization of approximate pairwise matchings based on the assignment of adjacency matrix eigenvectors, which correspond in near isomorphic molecular graphs. We take advantage of the fact that the minimal spanning set of best bipartite assignments fully describes the multipartite matching, which is recovered via its transitive closure. Symmetries that are not relevant within the scope of the training dataset are successfully ignored. (D) This enables the efficient construction of individual kernel functions for each training molecule, reflecting the joined similarity of all its symmetric variants with another molecule. The kernel exercises the symmetries by consolidating all training examples in an arbitrary reference configuration from which they are distributed across all symmetric subdomains. This approach effectively trains the fully symmetrized dataset without incurring the additional computational cost.}
\label{fig:overview}
\end{figure*}
One possible solution to the accuracy problem is provided by direct \textit{ab initio} molecular dynamics (AIMD) simulations, where the quantum-mechanical forces are computed on the fly for atomic configurations at every time step~\cite{Tuckerman2010}. The majority of AIMD simulations employ the current workhorse method of electronic-structure theory, namely density-functional approximations (DFA) to the exact solution of the Schr\"odinger equation for a system of nuclei and electrons. Unfortunately, different DFAs yield contrasting results~\cite{Koch-Holthausen} for the structure, dynamics, and properties of molecular systems. Furthermore, DFA calculations are not systematically improvable. Alternatively, explicitly correlated methods beyond DFA could also be used in AIMD simulations, unfortunately this leads to a steep increase in the required computational resources, for example a nanosecond-long MD trajectory for a single ethanol molecule executed with the CCSD(T) method would take roughly a million CPU years on modern hardware. An alternative is a direct fit of the potential-energy surface (PES) from a large number of CCSD(T) calculations, however this is only practically achievable for rather small and rigid molecules~\cite{Partridge-Schwenke,Tew2014,marx1}.
 
To solve this accuracy and molecular size dilemma and furthermore to enable converged AIMD simulations close to the exact solution of the Schr\"odinger equation, here we develop an alternative approach using symmetrized gradient-domain machine learning (sGDML) to construct force fields with the accuracy of high-level \textit{ab initio} calculations. Recently, a wide range of sophisticated machine learning (ML) models for small molecules and elemental materials~\cite{Behler2007,Bartok2010,Behler2012,Rupp2012,Montavon2013a,Bartok2013,Hansen2013,Hansen2015,Rupp2015,Bartok2015_GAP,Ramprasad2015,Rupp2015,DeVita2015,eickenberg2018solid,Behler2016,Ceriotti2016,Brockherde2017,artrith2017efficient,Shapeev2017,Ceriotti2017,Glielmo2017,Gastegger2017,Schutt2017,yao2017many,dral2017structure,john2017many,huang1707dna,faber2017prediction,huan2017universal,schutt2017schnet,noe2018,glielmo2018efficient,zhang2018deep,lubbers2018hierarchical,tang2018atomistic,Grisafi2018,ryczko2018convolutional,kanamori2018exploring,pronobis2018many,hy2018predicting,Smith2018,yao2018tensormol} have been proposed for constructing PES from DFA calculations. While these results are encouraging, direct ML fitting of molecular PESs relies on the availability of large reference datasets to obtain an accurate model. Frequently, those ML models are trained on thousands or even millions of atomic configurations. This prevents the construction of ML models using high-level \emph{ab initio} methods, for which energies and forces only for 100s of conformations can be practically computed. 

Instead, we propose a solution that allows converged molecular dynamics simulations with fully quantized electrons and nuclei for molecules with up to a few dozen atoms. This is enabled by two novel aspects: a reduction of the problem complexity through a data-driven discovery of relevant spatial and temporal physical symmetries, and enhancing the information content of data samples by exercising these identified static and dynamic symmetries, hence implicitly increasing the amount of training data.
Using the proposed sGDML approach, we carry out MD simulations at the \textit{ab initio} coupled cluster level of electronic-structure theory and provide insights into their dynamical behavior. Our approach contributes the key missing ingredient for achieving spectroscopic accuracy and rigorous dynamical insights in molecular simulations.

\section{Results}

\subsection{Symmetrized gradient-domain machine learning}
The sGDML model is built on the previously introduced GDML model~\cite{gdml}, but now incorporates all relevant physical symmetries, hence enabling MD simulations with high-level \emph{ab initio} force field accuracy. One can classify physical symmetries of molecular systems into symmetries of space and time and specific static and dynamic symmetries of a given molecule (see Fig.~\ref{fig:overview}). Global spatial symmetries include rotational and translational invariance of the energy, while homogeneity of time implies energy conservation. These global symmetries were already successfully incorporated into the GDML model~\cite{gdml}. Additionally, molecules possess well-defined rigid space group symmetries (i.e. reflection operation), as well as dynamic non-rigid symmetries (i.e. methyl group rotations). For example, the benzene molecule with only six carbon and six hydrogen atoms can already be indexed in $6!6! = 518400$ different, but physically equivalent ways. However, not all of these symmetric variants are accessible without crossing impassable energy barriers. Only the 24 symmetry elements in the $D_{6\text{h}}$ point group of this molecule are relevant. While methods for identifying molecular point groups for polyatomic rigid molecules are readily available~\cite{wilson1955molecular}, Longuet-Higgins~\cite{Longuet-Higgins1963} has pointed out that non-rigid molecules have extra symmetries. These dynamical symmetries arise upon functional-group rotations or torsional displacements and they are usually not incorporated in traditional force fields and electronic-structure calculations. Typically, extracting nonrigid symmetries requires chemical and physical intuition about the system at hand. Here we develop a physically-motivated algorithm for data-driven discovery of all relevant molecular symmetries from MD trajectories.

Molecular dynamics trajectories consist of smooth consecutive changes
in nearly isomorphic molecular graphs. When sampling from these trajectories
the combinatorial challenge is to correctly identify the same atoms
across the examples such that the learning method can use consistent
information for comparing two molecular conformations in its kernel
function. While so-called bi-partite matching allows to locally assign
atoms $\mat{R}= (\vec{r}_1, \dots, \vec{r}_N)$ for
each pair of molecules in the training set, this strategy alone is not
sufficient as it needs to be made globally consistent by multi-partite
matching in a second step.~\cite{Pachauri2013,Schiavinato2015,Kriege2016}

We start with adjacency matrices as representation for the
molecular graph~\cite{graph_kernels,Rupp2012,Hansen2015,Ferre2016,gdml}. To
solve the pairwise matching problem we therefore seek to find the
assignment $\tau$ which minimizes the squared Euclidean distance
between the adjacency matrices $\mat{A}$ of two isomorphic graphs $G$
and $H$ with entries $(\mat{A})_{ij} = \|\vec{r}_i - \vec{r}_j\|$,
where $\mat{P}(\tau)$ is the permutation matrix that realizes the
assignment:
\begin{equation}
\operatorname*{arg\,min}_{\tau} \mathcal{L}(\tau) = \|\mat{P}(\tau)\mat{A}_G\mat{P}(\tau)\tran - \mat{A}_H\|^2 \text{.}
\label{eq:matching_objective}
\end{equation}
Adjacency matrices of isomorphic graphs have identical eigenvalues and
eigenvectors, only their assignment differs. Following the approach of
Umeyama~\cite{Umeyama1988}, we identify the correspondence of
eigenvectors $\mat{U}$ by projecting both sets $\mat{U}_G$ and
$\mat{U}_H$ onto each other to find the best overlap. We use the overlap matrix, after sorting eigenvalues and overcoming sign ambiguity
\begin{equation}
\mat{M} = \text{abs}(\mat{U}_G) \text{abs}(\mat{U}_H)\tran \text{,}
\label{eq:cost_matrix}
\end{equation}
Then $-\mat{M}$ is provided as the cost matrix for the Hungarian
algorithm~\cite{Kuhn1955}, maximizing the overall overlap which
finally returns the approximate assignment $\tilde{\tau}$ that minimizes
Eq.~\ref{eq:matching_objective} and thus provides the results of step
one of the procedure.
As indicated, global inconsistencies may arise, e.g. violations of the
transitivity property $\tau_{jk} \circ \tau_{ij} = \tau_{ik}$ of the
assignments, therefore a second step is necessary which is based on
the composite matrix $\tilde{\mathcal{P}}$ of all pairwise assignment matrices
$\tilde{\mat{P}}_{ij} \equiv \mat{P}(\tilde{\tau}_{ij})$ within the
training set.

We propose to reconstruct a rank-limited $\mathcal{P}$ via the
transitive closure of the minimum spanning tree (MST) that minimizes
the bi-partite matching cost (see Eq.~\ref{eq:matching_objective},
Fig.~\ref{fig:overview}) over the training set. The MST is constructed
from the most confident bi-partite assignments and represents the rank
$N$ skeleton of $\tilde{\mathcal{P}}$, defining also $\mathcal{P}$.

The resulting {\em consistent multi-partite matching} $\mathcal{P}$ enables us to construct symmetric kernel-based ML models of the form
\begin{equation}
\hat{f}(\vec{x}) = \sum^M_{ij} \alpha_{ij} \kappa(\vec{x},\mat{P}_{ij} \vec{x}_i) \text{,}
\label{eq:data_augmentation_model}
\end{equation}
by augmenting the training set with the symmetric variations of each
molecule (see Supplementary Note 1 for a comparison with alternative symmetry-adapted kernel functions). A particular advantage of our solution is that it can fully populate all recovered permutational configurations even if they do not form a symmetric group, severely reducing the computational effort in evaluating the model.
Even if we limit the range of $j$ to include all $S$ unique
assignments only, the major downside of this approach is that a
multiplication of the training set size leads to a drastic increase in
the complexity of the cubically scaling kernel ridge regression
learning algorithm.
We overcome this drawback by exploiting the fact that the set of coefficients $\vec{\alpha}$ for the symmetrized training set exhibits the same symmetries as the data, hence the linear system can be contracted to its original size, while still defining the full set of coefficients exactly.

For notational convenience we transform all training geometries into a canonical permutation $\vec{x}_i \equiv \mat{P}_{i1}\vec{x}_i$, enabling the use of uniform symmetry transformations $\mat{P}_{j} \equiv \mat{P}_{1j}$ (see Supplementary Note 2). Simplifying Eq.~\ref{eq:data_augmentation_model} accordingly, gives rise to the symmetric kernel that we originally set off to construct
\begin{equation}
\begin{aligned}
\hat{f}(\vec{x}) &= \sum^M_i \alpha_{i} \sum^S_q  \kappa(\vec{x},\mat{P}_{q} \vec{x}_i) \\
&= \sum_i \alpha_{i} \kappa_\text{sym}(\vec{x},\vec{x}_i) \text{,}
\end{aligned}
\end{equation}
and yields a model with the exact same number of parameters as the original, non-symmetric one.

Our symmetric kernel is an extension to regular kernels and can be
applied universally, in particular to kernel based force fields. Here
we construct a symmetric variant of the gradient domain learning
(GDML) model, sGDML. This symmetrized GDML force field kernel takes
the form:
\begin{equation}
\text{Hess}(\kappa_{\text{sym}})(\vec{x},\vec{x}^{\prime}) = \sum^S_{q} \text{Hess}(\kappa)(\vec{x},\mat{P}_{q}\vec{x}^{\prime})\mat{P}_q \text{.}
\end{equation}
Accordingly, the trained force field estimator collects the contributions of the partial derivatives $3N$ of all training points $M$ and number of symmetry transformations $S$ to compile the prediction for a new input $\vec{x}$. It takes the form
\begin{equation}
\mat{\hat{f}_F}(\vec{x}) = \sum^M_{i} \sum^{3N}_{l} \sum^{S}_{q} (\mat{P}_q \vec{\alpha}_{i})_l  \frac{\partial}{\partial x_{l}} \nabla \kappa(\vec{x},\mat{P}_{q}\vec{x}_i)
\label{eq:force_model}
\end{equation}
and a corresponding energy predictor is obtained by integrating $\mat{\hat{f}_F}$ with respect to the Cartesian geometry. Due to linearity of integration, the expression for the energy predictor is identical up to second derivative operator on the kernel function.

Every (s)GDML model is trained on a set of reference examples that reflects the population of energy states a particular molecule visits during an MD simulation at a certain temperature. For our purposes, the corresponding set of geometries is subsampled from a 200 picosecond DFT MD trajectory at 500 K following the Boltzmann distribution. Subsequently, a globally consistent permutation graph is constructed that jointly assigns all geometries in the training set, providing a small selection of physically feasible transformations that define the training set specific symmetric kernel function. 
In the interest of computational tractability, we shortcut this sampling process to construct sGDML$@$CCSD(T) and only recompute energy and force labels at this higher level of theory.

The sGDML model can be trained in closed form, which is both quicker and more accurate than numerical solutions. Model selection is performed through a grid search on a suitable subset of the hyper-parameter space. Throughout, cross-validation with dedicated datasets for training, testing and validation are used to estimate the generalization performance of the model.

\begin{figure}[]
\centering
\includegraphics[width=\columnwidth]{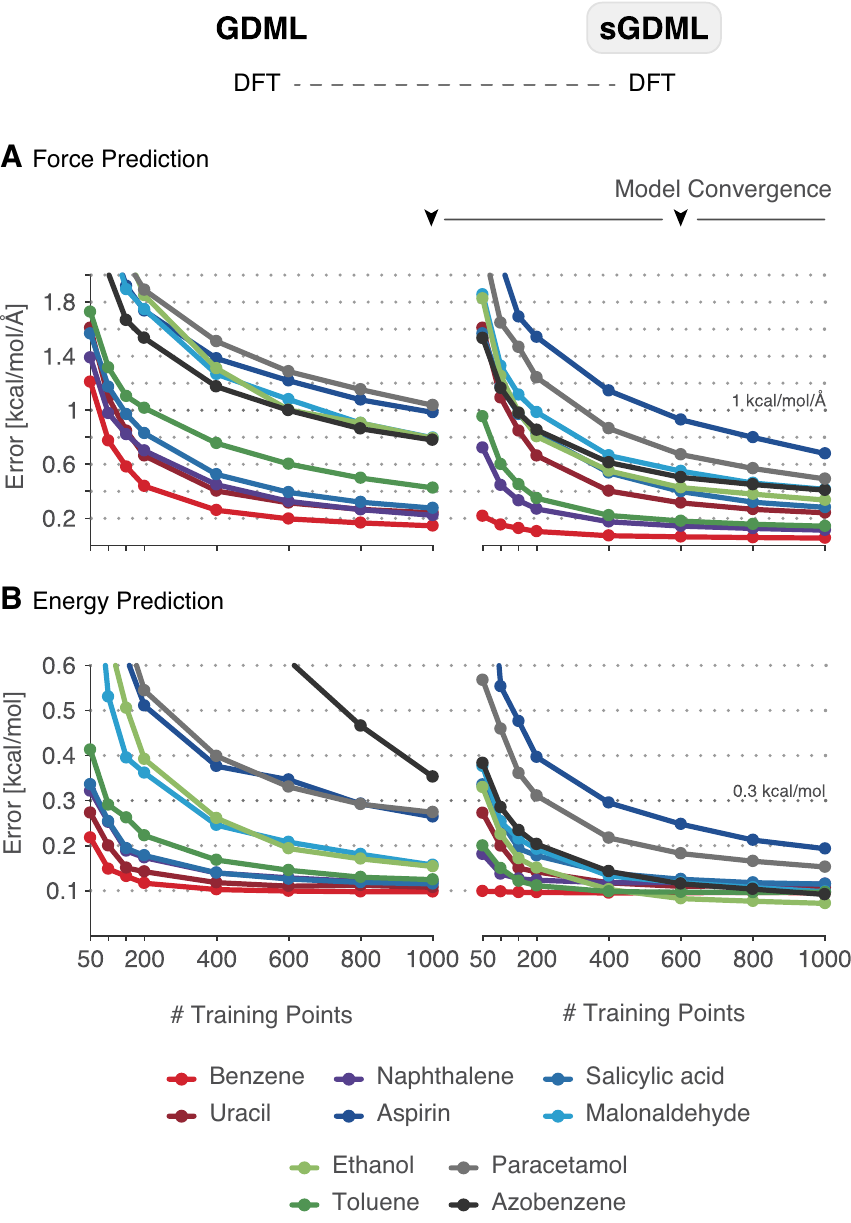}
\caption{Data efficiency gains using sGDML versus GDML. Energy and force prediction accuracy (in terms of the mean absolute error (MAE)) as a function of training set size of both models trained on DFT forces: the gain in efficiency and accuracy is directly linked to the number of symmetries in the system.}
\label{fig:learning_curves_gdml_cs_sgdml}
\end{figure}

\subsection{Forces and energies from GDML to \lowercase{s}GDML$@$DFT to \lowercase{s}GDML$@$CCSD(T)}

Our goal is to demonstrate that it is possible to construct compact sGDML models that faithfully recover CCSD(T) force fields for flexible molecules with up to 20 atoms, by using only a small set of few hundred molecular conformations. As a first step, we investigate the gain in efficiency and accuracy of sGDML model vs. GDML model employing MD trajectories of ten molecules from benzene to azobenzene computed with DFT (see Fig.~\ref{fig:learning_curves_gdml_cs_sgdml} and Supplementary Table 1). The benefit of a symmetric model is directly linked to the number of symmetries in the system. For toluene, naphthalene, aspirin, malonaldehyde, ethanol, paracetamol and azobenzene, sGDML improves the force prediction by $31.3\%$ to $67.4\%$ using the same training set in all cases (see Table~\ref{tab:sgdml_relative_performance}). As expected, uracil and salicylic acid have no exploitable symmetries, hence the performance of sGDML is unchanged with respect to GDML.
The inclusion of symmetries leads to a stronger improvement in force prediction performance compared to energy predictions. This is most clearly visible for the naphthalene dataset, where the force predictions even improve unilaterally. We attribute this to the difference in complexity of both quantities and the fact that an energy penalty is intentionally omitted in the cost function to avoid a tradeoff.
\begin{table}[]
  \centering
  \caption{Relative increase in accuracy of the sGDML$@$DFT vs. the non-symmetric GDML model: the benefit of a symmetric model is directly linked to the number of permutational symmetries in the system. All symmetry counts include the identity transformation.}
  \label{tab:sgdml_relative_performance}
  \setlength\extrarowheight{3pt}
  \begin{ruledtabular}
  \begin{tabular}{lrrr}
    Molecule & \# Sym. in $\kappa_\text{sym}$ & \multicolumn{2}{l}{$\Delta$ MAE [$\%$]}\\[0.9ex]
    \cline{3-4}
    & & Energy & Forces \\[0.9ex]
    \hline
    Benzene & 12 & -1.6 & -62.3\\
    Uracil & 1 & 0.0 & 0.0\\
    Naphthalene & 4 & 0.0 & -52.2\\
    Aspirin & 6 & -29.6 & -31.3\\
    Salicylic acid & 1 & 0.0 & 0.0\\
    Malonaldehyde & 4 & -37.5 & -48.8\\
    Ethanol & 6 & -53.4 & -58.2\\
    Toluene & 12 & -16.7 & -67.4\\
    Paracetamol & 12 & -40.7 & -52.9\\
    Azobenzene & 8 & -74.3 & -47.4\\
  \end{tabular}
  \end{ruledtabular}
\end{table}

A minimal force accuracy required for reliable MD simulations is $\text{MAE}=1$ kcal $\text{mol}^{-1} \text{\AA}^{-1}$. While the GDML model can achieve this accuracy at around 800 training examples for all molecules except aspirin, sGDML only needs 200 training examples to reach the same quality. Note that energy-based ML approaches typically require two to three orders of magnitude more data~\cite{gdml}. 

Given that the novel sGDML model is data efficient and highly accurate, we are now in position to tackle CCSD(T) level of accuracy with modest computational resources. We have trained sGDML models on CCSD(T) forces for benzene, toluene, ethanol, and malonaldehyde. For the larger aspirin molecule, we used CCSD forces (see Supplementary Table 2). The sGDML$@$CCSD(T) model achieves a high accuracy for energies, reducing the prediction error of sGDML$@$DFT by a factor of $1.4$ (for ethanol) to $3.4$ (for toluene). This finding leads to an interesting hypothesis that sophisticated quantum-mechanical force fields are smoother and, as a convenient side effect, easier to learn. Note that the accuracy of the force prediction in both sGDML$@$CCSD(T) and sGDML$@$DFT is comparable, with the benzene molecule as the only exception. We attribute this aspect to slight shifts in the locations of the minima on the PES between DFT and CCSD(T), which means that the data sampling process for CCSD(T) can be further improved. In principle, we can envision a corrected resampling procedure for CCSD(T), using the sGDML$@$CCSD(T) model as future work.

\subsection{Molecular dynamics with \emph{ab initio} accuracy}

\begin{figure*}[]
\centering
\includegraphics[width=\textwidth]{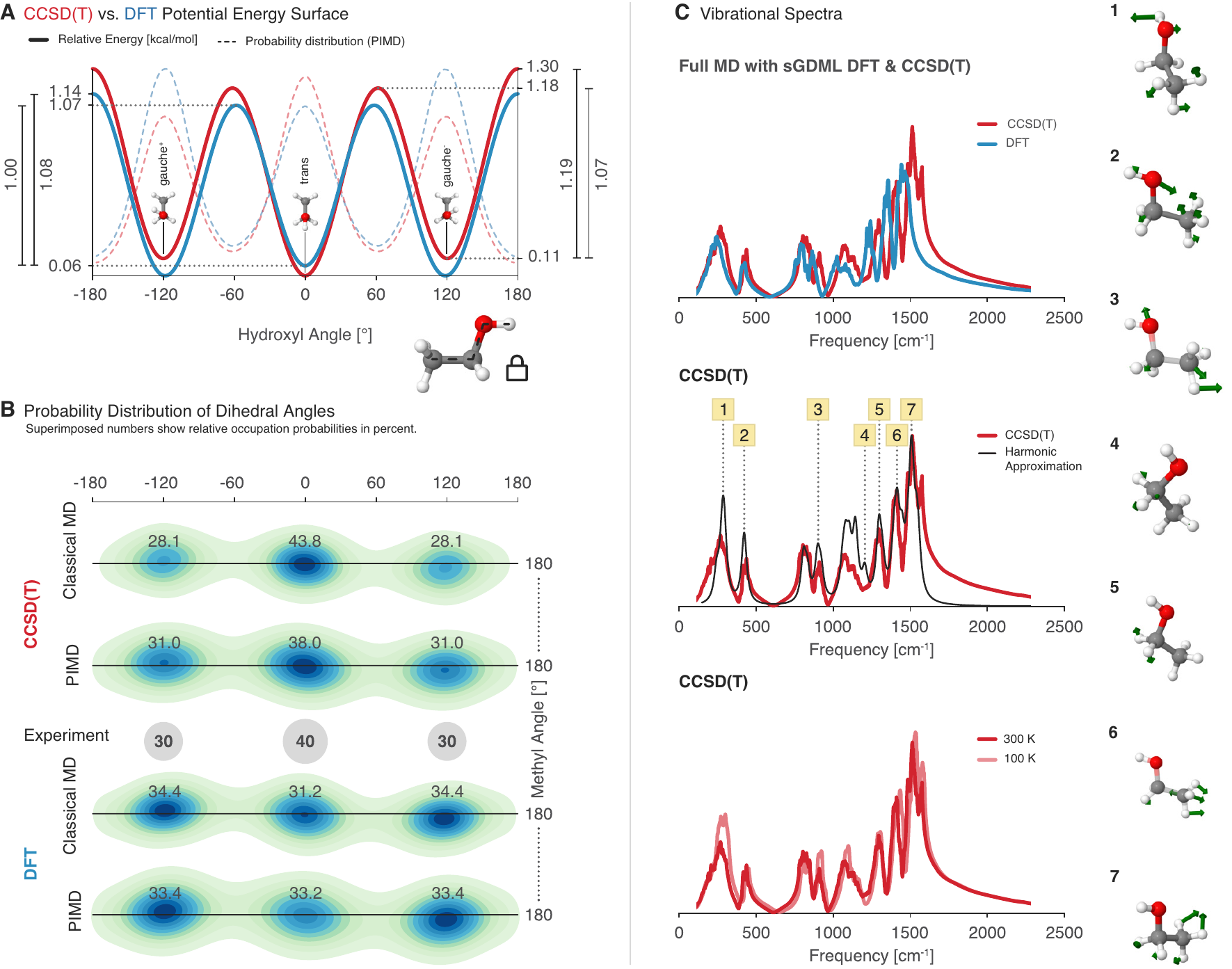}
\caption{Molecular dynamics simulations for ethanol. (A) Potential energy profile of the dihedral angle describing the rotation of the hydroxyl group for CCSD(T) (red) vs. DFT (blue). The energetic barriers predicted by sGDML@CCSD(T) are: $M_\text{t} \to M_\text{g}$: 1.18 kcal $\text{mol}^{-1}$, $M_\text{g-} \to M_\text{g+}$: 1.19 kcal $\text{mol}^{-1}$, and $M_\text{g}\to M_\text{t}$: 1.07 kcal $\text{mol}^{-1}$. The dashed lines show the probability distributions obtained from PIMD at 300K. (B) Joint probability distribution function for the two dihedral angles of the methyl and hydroxyl functional groups. Each minimum is annotated with the occupation probability obtained from classical and path-integral MD in comparison with experimental values. (C) Analysis of vibrational spectra (velocity--velocity autocorrelation function). (top) Comparison between the vibrational spectrum obtained from PIMD simulations at 300K for sGDML@CCSD(T) and its sGDML@DFT counterpart; (middle) comparison between the sGDML@CCSD(T) PIMD spectrum and the harmonic approximation based on CCSD(T) frequencies; (bottom) comparison of sGDML@CCSD(T) PIMD spectra at 300K and 100K. The rightmost panel shows several characteristic normal modes of ethanol, where atomic displacements are illustrated by green arrows.}
\label{fig:application_overview}
\end{figure*}

The predictive power of a force field can only be truly assessed by computing dynamical and thermodynamical observables, which require sufficient sampling of the configuration space, for example by employing molecular dynamics or Monte Carlo simulations. We remark that global error measures, such as mean average error (MAE) and root mean squared error (RMSE) are typically prone to overestimate the reconstruction quality of the force field, as they average out local topological properties. However, these local properties can become highly relevant when the model is used for an actual analysis of MD trajectories. 
As a demonstration, we will use the ethanol molecule; this molecule has three minima, $gauche\pm$ ($M_{\text{g}\pm}$) and $trans$ ($M_\text{t}$) shown in Fig.~\ref{fig:application_overview}-A, where experimentally it has been confirmed that $M_\text{t}$ is the ground state and $M_\text{g}$ is a local minimum~\cite{gonzalez1999}. The energy difference between these two minima is only 0.12 kcal $\text{mol}^{-1}$ and they are separated by an energy barrier of 1.15 kcal $\text{mol}^{-1}$. Obviously, the widely discussed ML target accuracy of 1 kcal $\text{mol}^{-1}$ is not sufficient to describe the dynamics of ethanol and other molecules.

This brings us to another crucial issue for predictive models: the reference data accuracy. Computing the energy difference between $M_\text{t}$ and $M_\text{g}$ using DFT(PBE-TS) we observe that $M_\text{g}$ is 0.08 kcal $\text{mol}^{-1}$ more stable than $M_\text{t}$, contradicting the experimental measurements. Repeating the same calculation using CCSD(T)/cc-pVTZ we find that $M_\text{t}$ is more stable than $M_\text{g}$ by 0.08 kcal $\text{mol}^{-1}$, in excellent agreement with experiment. From this analysis and subsequent MD simulations we conclude that CCSD(T) or sometimes even higher accuracy is necessary for truly predictive insights.

Additionally to requiring highly accurate quantum chemical approximations, the ethanol molecule also belongs to a category of fluxional molecules sensitive to nuclear quantum effects (NQE). This is because internal rotational barriers of the ethanol molecule ($M_\text{g}\leftrightarrow M_\text{t}$) are on the order of $\sim$1.2 kcal $\text{mol}^{-1}$ (see Fig.~\ref{fig:application_overview}), which is neither low enough to generate frequent transitions nor high enough to avoid them. In a classical MD at room temperature the thermal fluctuations lead to inadequate sampling of the PES. By correctly including NQE via path-integral molecular dynamics (PIMD), the ethanol molecule is able to transition between $M_\text{g}$ and $M_\text{t}$ configurations, radically increasing the transition frequency (see Supplementary Figure 1) and generating statistical weights in excellent agreement with experiment. Fig.~\ref{fig:application_overview}-B shows the statistical occupations of the different minima for ethanol using classical MD and PIMD for the sGDML@CCSD(T) and sGDML@DFT models in comparison with the experimental results. Overall, our MD results for ethanol highlight the necessity of using a highly accurate force field with an equally accurate treatment of NQE for achieving reliable and quantitative understanding of molecular systems.

Having established the accuracy of statistical occupations of different states of ethanol, we are now in position to discuss for the first time the CCSD(T) vibrational spectrum of ethanol computed using the velocity--velocity autocorrelation function based on centroid PIMD (see Fig.~\ref{fig:application_overview}-C). As a reference, in Fig.~\ref{fig:application_overview}-C-top we compare the vibrational spectra from DFT and CCSD(T) sGDML models in the fingerprint zone, and as expected the sGDML@CCSD(T) model generates higher frequencies but both share similar shapes but slightly different peak intensities. Molecular vibrational spectra at finite temperature include anharmonic effects, hence anharmonicities can be studied by comparing the sGDML@CCSD(T) spectrum with the harmonic approximation. Fig.~\ref{fig:application_overview}-C-middle shows such comparison and demonstrates that low-frequency and non-symmetric vibrations are most affected by finite-temperature contributions. The thermal frequency shift can be better seen in Fig.~\ref{fig:application_overview}-C-bottom, where the sGDML@CCSD(T) spectrum is compared at two different temperatures. We observe that each normal mode is shifted in a specific manner and not by a simple scaling factor, as typically assumed. The most striking finding from our simulations is the resolution of the apparent mismatch between theory and experiment explaining the origin of the torsional frequency for the hydroxyl group. Experimentally, the low frequency region of ethanol, around $\sim$210 cm$^{-1}$, is not fully understood, but there are frequency measurements for the hydroxyl rotor ranging in between $\sim$202~\cite{durig1990torsional,wassermann2010ethanol} and $\sim$207~\cite{durig1975raman} cm$^{-1}$ for gas-phase ethanol, while theoretically we found 243.7 cm$^{-1}$ at the sGDML@CCSD(T) level of theory in the harmonic approximation. From the middle and bottom panels in Fig.~\ref{fig:application_overview}-C, we observe that by increasing the temperature the lowest peak shifts to substantially lower frequencies compared to the rest of the spectrum. The origin of such phenomena is the strong anharmonic behavior of the lowest normal mode \textit{a}, shown in Fig.~\ref{fig:application_overview}-C-middle, which mainly corresponds to hydroxyl group rotations. At room temperature the frequency of this mode drops to $\sim$215 cm$^{-1}$, corresponding to a red-shift of 12\% and getting closer to the experimental results demonstrating the importance of dynamical anharmonicities. 

\begin{figure*}
\centering
\includegraphics[width=\textwidth]{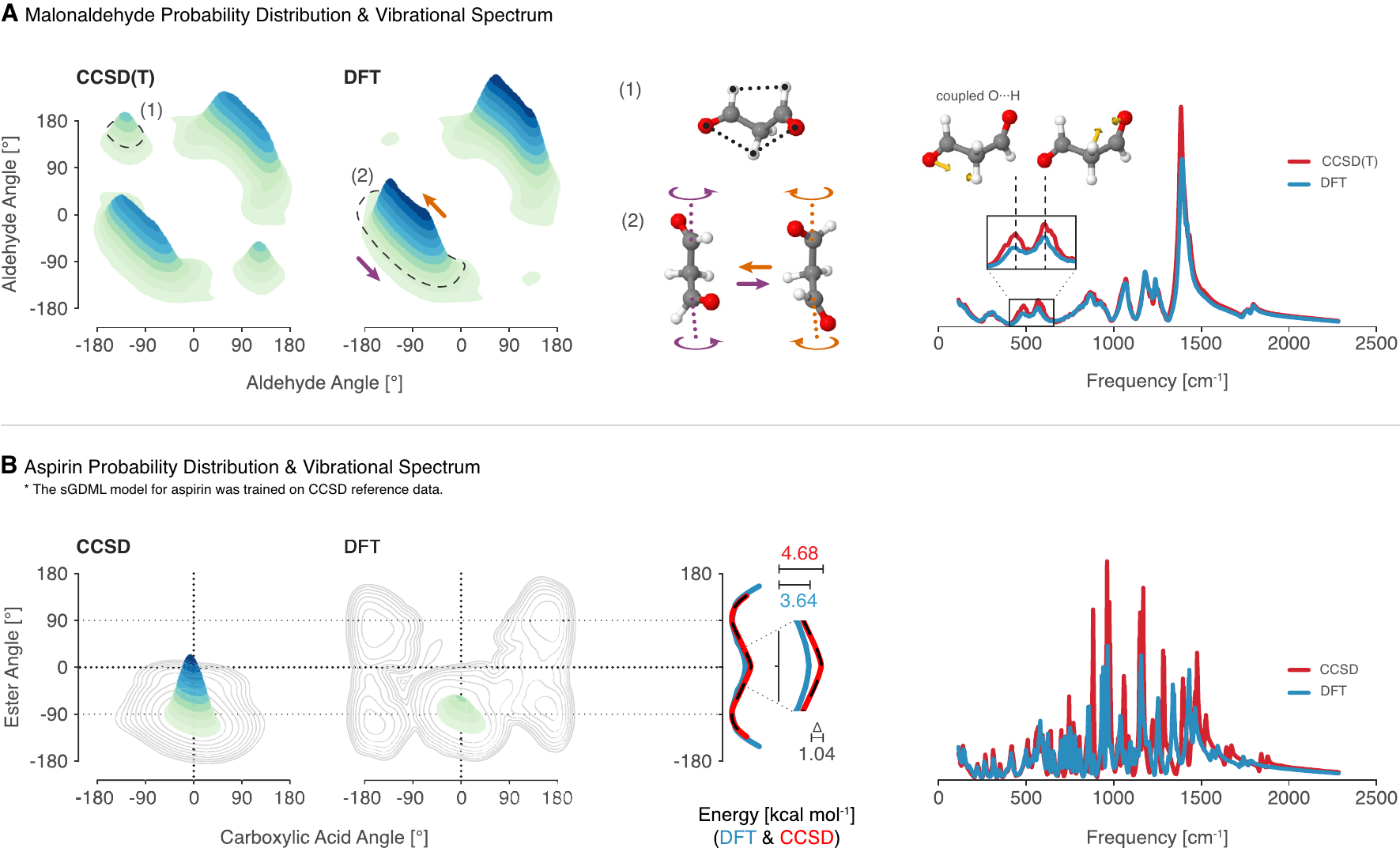}
\caption{Analysis of MD simulations with sGDML for malonaldehyde and aspirin. The MD simulations at 300 K were carried out for 500 ps. (A) Joint probability distributions of the dihedral angles in malonaldehyde, describing the rotation of both aldehyde groups based on classical MD simulations for sGDML@CCSD(T) and sGDML@DFT. The configurations (1) and (2) are representative structures of the most sampled regions of the PES. (B) Joint probability distributions of the dihedral angles in aspirin, describing the rotation of the ester and carboxylic acid groups based on PIMD simulations for sGDML@CCSD and sGDML@DFT using 16 beads at 300 K. The potential energy profile for the ester angle in kcal $\text{mol}^{-1}$ is shown for sGDML@CCSD (red), sGDML@DFT (blue) and compared with the CCSD reference (black, dashed). Contour lines show the differences of both distributions on a log scale. Both panels also show a comparison of the vibrational spectra generated via the velocity-velocity autocorrelation function obtained with sGDML@CCSD(T)/CCSD (red) and sGDML@DFT (blue).}
\label{fig:application_overview_two}
\end{figure*}

Finally, we illustrate the wider applicability of the sGDML model to more complex molecules than ethanol by performing a detailed analysis of MD simulations for malonaldehyde and aspirin. In Fig.~\ref{fig:application_overview_two}-A, we show the joint probability distributions of the dihedral angles (PDDA) for the malonaldehyde molecule. This molecule has a peculiar PES with two local minima with a O$\cdots$H$\cdots$O symmetric interaction (structure (1)), and a shallow region where the molecule fluctuates between two symmetric global minima (structure (2)). The dynamical behavior represented in structure (2) is due to the interplay of two molecular states dominated by an intramolecular O$\cdots$H interaction and a low crossing barrier of $\sim$0.2 kcal $\text{mol}^{-1}$. An interesting result is the nearly unvisited structure (1) by sGDML@DFT in comparison to sGDML@CCSD(T) model regardless of the great similarities of their PES, which gives an idea of the observable consequences of subtle energy differences in the PES of molecules with several degrees of freedom. In terms of spectroscopic differences, the two approximations generate spectra with very few differences (Fig.~\ref{fig:application_overview_two}-A-right), but being the most prominent the one between the two peaks around 500 cm$^{-1}$. Such difference can be traced back to the enhanced sampling of the structure (1), and additionally it could be associated to the different nature between the methods in describing the intramolecular O$\cdots$H coupling.

For aspirin, the consequences of proper inclusion of the electron correlation are even more significant. Fig.~\ref{fig:application_overview_two}-B shows the PIMD generated PDDA for DFT and CCSD based models. By comparing the two distributions we find that sGDML@CCSD generates localized dynamics in the global energy minimum, whereas the DFT model yields a rather delocalized sampling of the PES. These two contrasting results are explained by the difference in the energetic barriers along the ester dihedral angle. The incorporation of electron correlation in CCSD increases the internal barriers by $\sim$1 kcal $\text{mol}^{-1}$. This prediction was corroborated with explicit CCSD(T) calculations along the dihedral-angle coordinate (black dashed line in Fig.~\ref{fig:application_overview_two}-B-PES). Furthermore, the difference in the sampling is also due to the fact that the DFT model generates consistently softer interatomic interactions compared to CCSD, which leads to large and visible differences in the vibrational spectra between DFT and CCSD (Fig.~\ref{fig:application_overview_two}-B-right). 

\section{Discussion}

\begin{figure}[]
\centering
\includegraphics[width=\columnwidth]{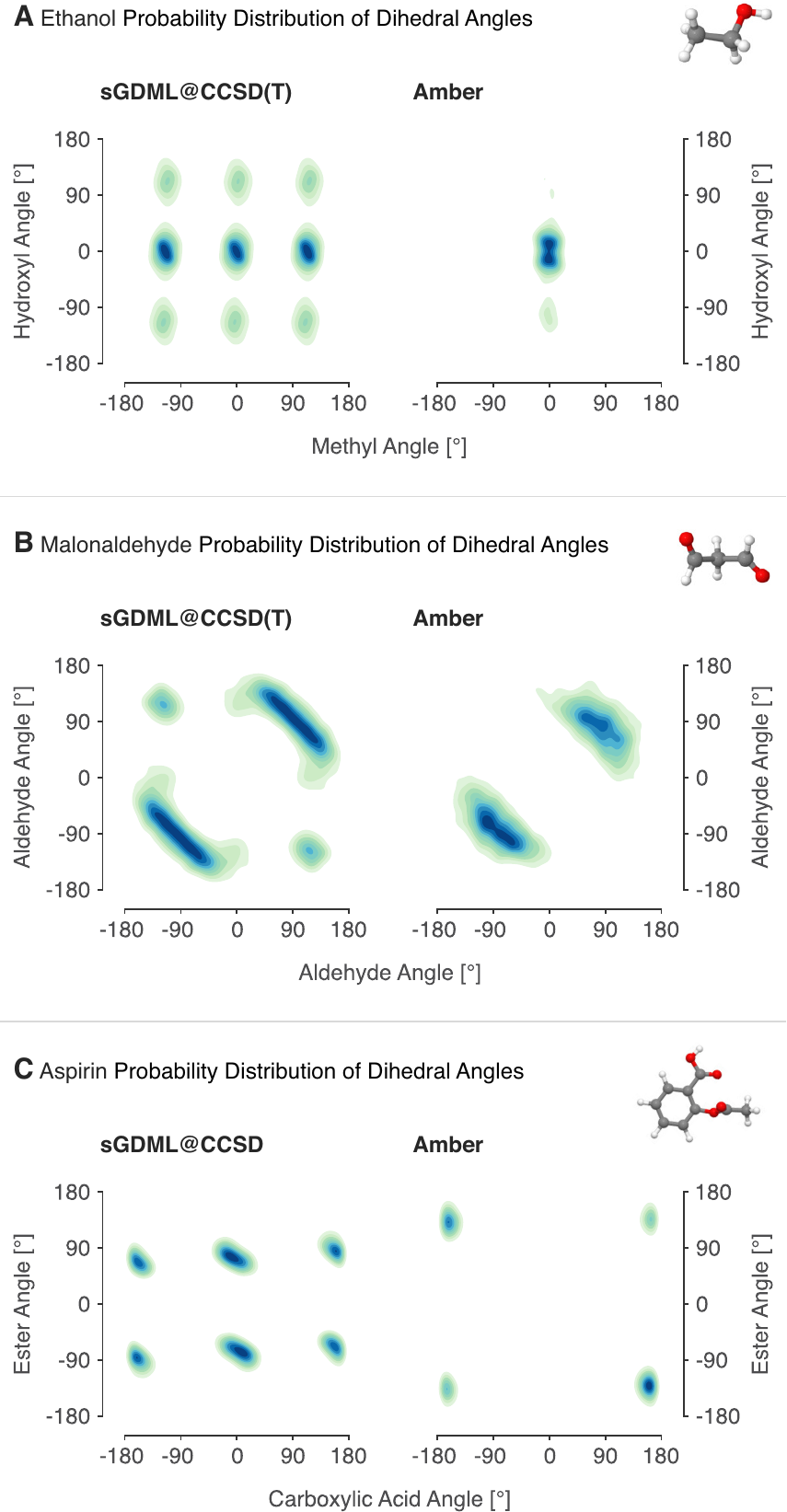}
\caption{Accuracy of the sGDML model in comparison to a traditional force field. We contrast the dihedral angle probability distributions of ethanol, malonaldehyde and aspirin obtained from classical MD simulations at 300 K with sGDML (left column) versus the AMBER~\cite{amber18} (right column) force field. The ethanol simulations were carried out at constant energy (NVE), whereas a constant temperature (NVT) was used for malonaldehyde and aspirin. (A) Ethanol: the coupling between the hydroxyl and methyl rotor is absent in AMBER. Moreover, the probability distribution shows an unphysical harmonic sampling at room temperature, revealing the oversimplified harmonic description of bonded interactions in that force field. (B, C) Malonaldehyde and aspirin: the formulation of the AMBER force field is dominated by Coulomb interactions, which can lead an incomplete description of the PES and even spurious global minima in the case of aspirin. The length of the simulations was 0.5 ns.}
\label{fig:forcefield_comparison}
\end{figure}

The present work enables molecular dynamics simulations of flexible molecules with up to a few dozen atoms with the accuracy of high-level \textit{ab initio} quantum mechanics. Such simulations pave the way to computations of dynamical and thermodynamical properties of molecules with an essentially exact description of the underlying potential-energy surface. On the one hand, this is a required step towards molecular simulations with spectroscopic accuracy. On the other, our accurate and efficient sGDML model leads to unprecedented insights when interpreting the experimental vibrational spectra and dynamical behavior of molecules. The contrasting demands of accuracy and efficiency are satisfied by the sGDML model through a rigorous incorporation of physical symmetries (spatial, temporal, and local symmetries) into a gradient-domain machine learning approach. This is a significant improvement over symmetry adaption in traditional force fields and electronic-structure calculations, where usually only (global) point groups are considered. Global symmetries are increasingly less likely to occur with growing molecule size, providing diminishing returns. Local symmetries on the other hand are system size independent and preserved even when the molecule is fragmented for large-scale modeling.

In many of the applications of machine learned force fields the target error is the chemical accuracy or thermochemical accuracy (1 kcal $\text{mol}^{-1}$), but this value was conceived in the sense of thermochemical experimental measurements, such as heats of formation or ionization potentials. Consequently, the accuracy in ML models for predicting the molecular PES should not be tied to this value. Here, we propose a framework for the accuracy in force fields which satisfy the stringent demands of molecular spectroscopists, being typically in the range of wavenumbers ( $\approx$ 0.03 kcal $\text{mol}^{-1}$). Reaching this accuracy will be one of the greatest challenges of ML-based force fields. We remark that energy differences between molecular conformers are often on the order of 0.1--0.2 kcal $\text{mol}^{-1}$, hence reaching spectroscopic accuracy in molecular simulations is needed to generate predictive results.

A comparable accuracy is not obtainable with traditional force fields (see Fig.~\ref{fig:forcefield_comparison}). 
In general, they miss most of the crucial quantum effects due to their rigid, handcrafted analytical form. For example, the absence of a term for electron lone pairs in AMBER leads to uncoupled rotors in ethanol. Furthermore the oversimplified harmonic description of bonded interactions generates an unphysical harmonic sampling at room temperature (see Fig.~\ref{fig:forcefield_comparison}-A). 
In the case of malonaldehyde (Fig.~\ref{fig:forcefield_comparison}-B), both distributions misleadingly resemble each other, however they emerge from different types of interactions. For AMBER, the dynamics are purely driven by Coulomb interactions, while the sampling with sGDML@CCSD(T) (structure (2) in Fig.~\ref{fig:application_overview_two}-A) is mostly guided by electron correlation effects. Lastly, a complete mismatch between the regular force field and sGDML is evident for aspirin (see Fig.~\ref{fig:forcefield_comparison}-C), where the interactions dominated by Coulomb forces generate a completely different PES with spurious global and local minima. It is worth mentioning, that the observed shortcomings of the AMBER force field can be addressed for a particular molecule, however only at the cost of losing generality and computational efficiency.

In the context of machine learning, our work connects to recent studies on the usage of invariance constraints for learning and representations in vision. In the human visual system and also in computer vision algorithms the incorporation of invariances such as translation, scaling and rotation of objects can in principle permit higher performance at more data efficiency~\cite{poggio2016visual}; learning theoretical bounds can furthermore show that the amount of data required is reduced by a factor: the number of parameters of the invariance transformation~\cite{anselmi2016invariance}. Interestingly, our study goes empirically beyond this factor, i.e.~our gain in data efficiency is often more than two orders of magnitude when combining the invariances (physical symmetries). We speculate that our finding may indicate that the learning problem itself may become less complex, i.e. that the underlying problem structure becomes significantly easier to represent.

There is a number of challenges that remain to be solved to extend the sGDML model in terms of its applicability and scaling to larger molecular systems. Given an extensive set of individually trained sGDML models, an unseen molecule can be represented as a non-linear combination of those models. This would allow scaling up and transferable prediction for molecules that are similar in size.
Advanced sampling strategies could be employed to combine forces from different levels of theory to minimize the need for computationally-intensive \textit{ab initio} calculations. Our focus in this work was on intramolecular forces in small- and medium-sized molecules. Looking ahead, it is sensible to integrate the sGDML model with an accurate intermolecular force field to enable predictive simulations of condensed molecular systems (Ref.~\cite{Tristan} presents an intermolecular model which would be particularly suited for coupling with sGDML). Many other avenues for further development exist~\cite{de2017use}, including incorporating additional physical priors, reducing dimensionality of complex PES, computing reaction pathways, and modeling infrared, Raman, and other spectroscopic measurements.

\section{Methods}

\subsection{Reference data generation}
The data used for training the DFT models were created running \textit{ab initio} MD in the NVT ensemble using the Nos\'e-Hoover thermostat at 500 K during a 200 ps simulation with a resolution of 0.5 fs. We computed forces and energies using all-electrons at the generalized gradient approximation (GGA) level of theory with the Perdew-Burke-Ernzerhof (PBE)~\cite{PBE1996} exchange-correlation functional, treating van der Waals interactions with the Tkatchenko-Scheffler (TS) method~\cite{TS}. All calculations were performed with FHI-aims~\cite{FHIaims2009}. The final training data was generated by subsampling the full trajectory under preservation of the Maxwell-Boltzmann distribution for the energies.

To create the coupled cluster datasets, we reused the same geometries as for the DFT models and recomputed energies and forces using all-electron
coupled cluster with single, double, and perturbative triple excitations (CCSD(T)). The Dunning's correlation-consistent basis set 
cc-pVTZ was used for ethanol, cc-pVDZ for toluene and malonaldehyde and CCSD/cc-pVDZ for aspirin. All calculations were performed with the Psi4~\cite{parrish2017psi4} software suite.

\subsection{Molecular dynamics}
In order to incorporate the crucial effects induced by quantum nuclear delocalization, we used path-integral molecular dynamics (PIMD) which incorporates quantum-mechanical effects into molecular dynamics simulations via the Feynman's path integral formalism. The PIMD simulations were performed with the sGDML model interfaced to the i-PI code~\cite{ceriotti2014pi}. The integration timestep was set to 0.2 fs to ensure energy conservation along the MD using the NVE and NVT ensemble. The total simulation time was 1 ns for ethanol (Fig. 3) to get a converged sampling of the PES using 16 beads in the PIMD.

\subsection{Bipartite matching cost matrix}
For the bipartite matching of a pair of molecular graphs, we solve the optimal assignment problem for the eigenvectors of their adjacency matrices using the Hungarian algorithm~\cite{Kuhn1955}. As input, this algorithm expects a matrix with all pairwise assignment costs $\mat{C}_{\mat{M}} =  -\mat{M} $, which is constructed as the negative overlap matrix from Eq. 2. We add a penalty matrix with entries $(\mat{C}_\vec{z})_{ij} = \text{abs}((\vec{z})_i - (\vec{z})_j)\epsilon$ that prevents the matching of non-identical nuclei for sufficiently large $\epsilon > 0$. The final const matrix is then $\mat{C} = \mat{C}_{\mat{M}} + \mat{C}_\vec{z}$.

\subsection{Training sGDML}

The symmetric kernel formulation approximates the similarities in the kernel matrix between different permutational configurations of the inputs, as they would appear with a fully symmetrized training set. We construct this object as the sum over all relevant atom assignments for each training geometry, such that the kernel matrix retains its original size. This procedure is used to symmetrize the GDML model~\cite{gdml}, where the symmetric kernel function takes the form
\begin{equation}
\text{Hess}(\kappa_{\text{sym}})(\vec{x},\vec{x}^{\prime}) = \frac{1}{S}\sum^S_{pq} \mat{P}\tran_p\text{Hess}(\kappa)(\mat{P}_p\vec{x},\mat{P}_{q}\vec{x}^{\prime})\mat{P}_q \text{.}
\end{equation}
Note, that the rows and columns of the Hessian in the summand are permuted (using $\mat{P}\tran_p$ and $\mat{P}_q$) such that the corresponding partial derivatives align. When evaluating the model, the free variable $\vec{x}$ (first argument of the kernel function) is not permuted and the normalization factor is dropped (see Eq. 5). See Supplementary Note 3 for information on how to use the sGDML model, when the input is represented by a descriptor.

\subsection{Data Availability}
All datasets used in this work are available at http://quantum-machine.org/datasets/. Additional data related to this paper may be requested from the authors.

\bibliographystyle{naturemag}
\bibliography{references/main.bbl}

\section*{Acknowledgements}
We thank Michael Gastegger for providing the AMBER force fields. S.C., A.T., and K.-R.M. thank the Deutsche Forschungsgemeinschaft (project MU 987/20-1) for funding this work. A.T. is funded by the European Research Council with ERC-CoG grant BeStMo. K.-R.M. gratefully acknowledges the BK21 program funded by the Korean National Research Foundation grant (no. 2012-005741). Part of this research was performed while the authors were visiting the Institute for Pure and Applied Mathematics, which is supported by the NSF.



\end{document}